\documentclass[aps,twocolumn,superscriptaddress,nofootinbib]{revtex4}
\usepackage{graphicx}
\usepackage{nicefrac}

\begin{document}


\title{Features of Kamiokande-II, IMB and Baksan observations  
and their interpretation in a two-component model for the signal}
\author{Francesco  Vissani}
\affiliation{INFN, Laboratori Nazionali del Gran Sasso, Assergi (AQ),  Italy}
\author{Giulia Pagliaroli}
\affiliation{University of L'Aquila, Coppito,  
and INFN, Laboratori Nazionali del Gran Sasso,  Italy}
\begin{abstract}
We consider the time, angular and energy
distributions of SN1987A events discussing 
the quality of their  agreement with the expectations. 
A global interpretation is performed 
considering a simple model, based on the standard scenario
for the explosion.
Despite the  contrasting and confusing indications, a straightforward fit 
to the data provides a result that does not contradict but rather supports
the expectations. 
The calculated electron antineutrino 
flux is applied to predict the relic neutrino signal.  
\end{abstract}
\maketitle

\section{Introduction}
The Symposium held in Moscow on February 2007 \cite{symp}
was devoted to discuss open problems of SN1987A. 
The meaning of LSD events~\cite{mb} was discussed, and  
the potential of the model of Imshennik and Ryazhskaya~\cite{ir} 
was shown. The observations of
Kamiokande-II \cite{kii}, IMB \cite{imb} and Baksan  \cite{bak}
were covered by the contribution of Alexeev.  
In the spirit of the Symposium, 
we recall that (1)~certain features of these observations 
are difficult to understand, see e.g.\ \cite{olgarev}; 
(2)~there is not yet a complete model for the explosion, 
that could shed doubts on the assumption that we  
know how neutrinos are emitted;  
this applies also to ref.~\cite{ir}.

In this work that expands the talk of F.V.\ at \cite{symp}
we study the features of Kamiokande-II, IMB and Baksan data 
and analyze them in the `standard scenario' 
of emission, first advocated by 
Nadyozhin \cite{nad}--also called 
Bethe and Wilson scenario \cite{bw},
delayed explosion or $\nu$-assisted explosion. 
We recall its features:
On top of a several-seconds lasting
emission from the proto neutron
star, the {\bf cooling}, there is 
an initial intense emission 
lasting a fraction of a second.  
The $\nu_e$ and $\bar\nu_e$
irradiated in this phase transfer energy and 
help the stalled shock wave to resume, 
eventually leading to explosion.   We call this phase of
neutrino emission {\bf accretion}. 
For 
progresses toward the implementation  of the `standard 
scenario' in numerical simulations, see~\cite{janka}.
 

\section{Features of the data\label{sec2}}
The observed events are 11 or 12 in Kamiokande-II, the $6^{th}$ 
being below the threshold of 7.5 MeV.
The duration and the energy of the signal  
is not safely known. Thus we consider a unified time
window of $T=30$~s in this work. 
The candidate signal events
increases to 16 in Kamiokande-II 
(the events number 13,14,15,16 being below 
the threshold), it is 8 in IMB and 5 in Baksan. 
Recall that most 
signal is due to 
$\bar\nu_e p\to e^+ n$ [IBD reaction]
that has the
largest cross section; unless stated otherwise we assume that all
signal events are due to IBD.

The background in Kamiokande-II declines rapidly
with the number of hit phototubes; on average, 
there should be about 0.7 background events 
above threshold and 5.6 in total.
Kamiokande-II analyzed the whole volume of the detector. The
region close to the walls (in particular the upper wall) 
is the less safe against the risk of background. Five 
of the 11 events are close to the walls, lying in the 
outermost 4\% of the volume of the detector; they include
the events number 3,4,10 of the dataset, close
to the threshold.
We expect 1 background event in Baksan in $T=30$~s; 
this, together with a typical expected signal 
of about 1.6 antineutrinos ($\mu=2.6$)
gives a reasonable Poisson chance
$\mu^n e^{-\mu}/n\mbox{!}=8$~\% when $n=5$. IMB instead can be 
assumed to be background free.

\begin{figure}[t]
$$\includegraphics[width=0.44\textwidth,
height=0.30\textwidth,
angle=0]{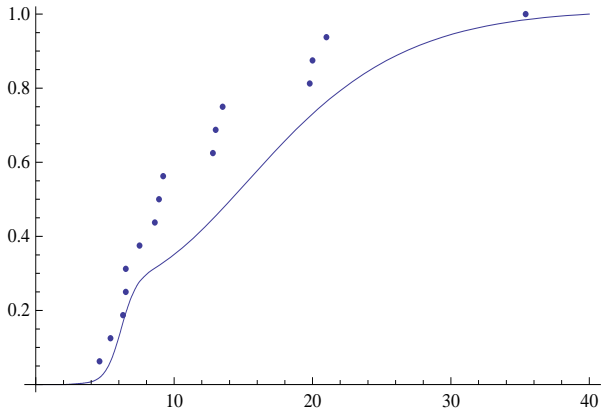}$$\vskip-4mm
\caption{\em \footnotesize 
Dots: observed energies in Kamiokande-II in 30 s. 
Continuous curve: expected cumulative 
spectrum (background + signal as in Eq.~\ref{appro}) 
as a function of the energy in MeV. 
The Smirnov-Cram\`e{}r-Von Mises (SCVM) \cite{scvm} goodness of fit is 19\%;
compare also with \cite{jcap}.\label{fig5}}
\end{figure}

\paragraph*{Energy distribution}
This distribution is difficult to interpret.
We select 3 questions for the discussion:\\
(i) Do the observed energies meet the expectations?\\
(ii) Are the average energies of IMB $31.9\pm 2.6$ MeV 
  and of Kamiokande-II $15.4\pm 2.4$ MeV compatible?\\
(iii) Why the first 4 Kamiokande-II events have average
  energy $12.6\pm 2.4$ MeV and the last seven $17.1\pm 3.3$ MeV?  \\
The first can be answered fitting the data with a black body (thermal)
$\bar\nu_e$ spectrum with luminosity $\propto R^2_c T^4_c$. We confirm that the
Kamiokande-II and IMB datasets are compatible only when we consider the
90\% C.L.; Baksan is compatible
with both. When we combine the data
in a unique analysis, the values of the
radius is larger than expected, $R_c\sim 30$ km.
The second question can be discussed quantitatively 
calculating the average energies of the events, keeping the
antineutrino temperature $T_c$ as a free parameter (see e.g.,
\cite{prd}):
Kamiokande-II and IMB suggest different values of $T_c$.   
The first two questions are the traditional formulations of the 
``energy problem''. 
They suggest two possible ways-out:
a)~The energy distribution is strongly non-thermal. This is explored,   
e.g., in \cite{mirizzi} and \cite{lunardini}.
Our objection is that this is a major deviation from the expectations 
that should not be admitted in a conservative analysis, as, {\em e.g.}, 
\cite{jh}. 
b)~Another possibility is that there are some 
low energy background events in Kamiokande-II~\cite{jcap}, 
see the feature at $E=4-8$~MeV in Fig.~\ref{fig5}, 
but recall that the background events above the nominal 
threshold are expected to be few.
The third formulation of the problem
\cite{ll},\cite{jcap} offers a new clue: the 
excess of low energy Kamiokande-II events is due 
to the early detected events. 
This is fine since the 
standard scenario predicts the existence
of the initial phase of accretion possibly with 
peculiar features. The high luminosity 
helps account the early events seen by Kamiokande-II.

\begin{figure}[b]
$$\includegraphics[width=0.40\textwidth,
height=0.36\textwidth,
angle=0]{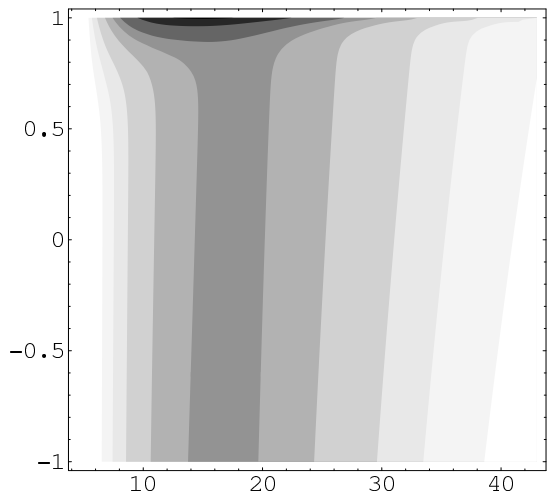}$$
\vskip-4mm
\caption{\em \footnotesize 
Expected 2-dimensional (positron-energy in MeV, cosine with the direction of 
SN1987A)-distribution for Kamiokande-II.
The contour plot with equally spaced heights shows  
the mild directionality of IBD reaction and the peak of 
ES events (forward region near 15 MeV) 
whose shape is dictated by instrumental effects (see e.g., \cite{prd}). 
\label{fa}}
\end{figure}

\paragraph*{Angular distribution}
The data  and the expected angular distribution 
have been recently compared in \cite{prd},  
 keeping into account the  angular bias of IMB \cite{imb}. The SCVM test shows
that the ``problem'' of the forward peak is not severe: 
if all event are IBD, the goodness of fit 
is 6.4\% for IMB and 8.6\% in Kamiokande-II. With 1 or 2 elastic 
scattering events (0.3-0.6 expected in Kamiokande-II) this improves
further. Note that, even being ready to consider  
something exotic, it
is hard to imagine a reaction that is forward-peaked but too much, 
as needed to locate half of the IMB events 
in the region $30^\circ <\theta<60^\circ$. 
Thus the discrepancy with the expectations
is not very compelling. 
We also estimated the angular
distribution taking into account elastic scattering events (ES). 
Assuming oscillations with normal mass hierarchy and 
increasing non-electronic neutrinos as allowed in 
\cite{keil}  we calculated the time-averaged 
distribution of Fig.~\ref{fa}.
From this, we find that the probability 
that the first event detected by Kamiokande-II is due to ES
is about 30\%.

\begin{figure}[t]
\vskip-4mm
$$\includegraphics[width=0.40\textwidth,
height=0.43\textwidth,
angle=270]{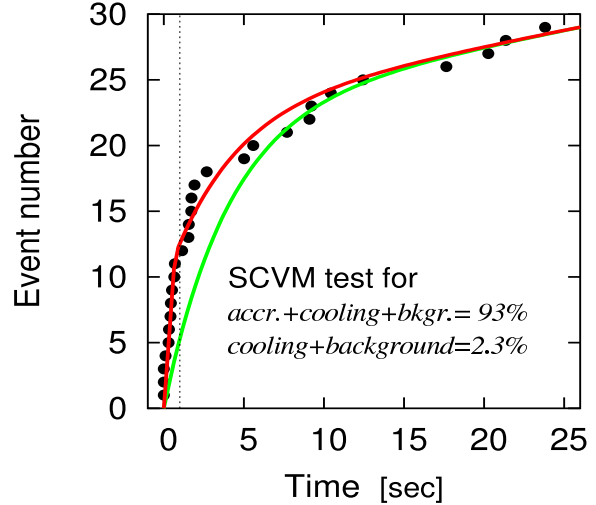}$$
\vskip-4mm
\caption{\em\footnotesize 
In the abscissa, the time of the event; in the 
ordinate, its progressive number. 
Two time distributions, comprising 
22 signal and 7 background events, 
are given. Also indicated their goodness of fit values.
In the lower one, all signal events belong to cooling, 
where the signal declines with a time constant $\tau_c=4$~s.
In the upper one,
only 13 signal events belong to cooling;
the remaining 9 belong to accretion phase and  
are distributed in the interval $t<0.7$~s.
The dotted vertical line marks the end of the first second.  
\label{ff}}
\end{figure}

\paragraph*{Time distribution}
The time sequence of the Kamiokande-II events is at first sight odd; 
there is a cluster of six events in the first second 
and a ``gap'' of 7 seconds between the event number 9 and the event
number 10, difficult to understand on physical basis.  
One way-out is to combine all data 
in a single dataset as in Fig.~\ref{ff}.
This synthetic dataset can be compared with the expectations. 
Besides the known background component, $n_b=6.6$ expected events, 
the signal should accumulate in two phases:
a first one with $n_a$ events, followed by 
a second and longer phase with $n_c=29-n_b-n_a$ events:
\begin{equation}
N(t)=n_b {\textstyle \frac{t}{T}} + 
n_a [{\textstyle 1\!+\!\theta(t\!-\!\tau_a) (
\frac{t}{\tau_a}-1)}] + 
n_c [1-e^{-{t}/{\tau_c}} ]
\end{equation}
where the suffixes $b,a,c$ mean background, accretion, cooling. 
The goodness-of-fit values shown in Fig.~\ref{ff} suggest 
that a two component model of this type has no 
problem to reproduce the data.\footnote{This 
still leaves some open questions:
The absolute times of Kamiokande-II and
Baksan are not reliable. Is it fair
to use the corresponding freedom 
of interpretation to set the first event in each detectors 
at $t=0$?
Do the assumed number of events fits into a reasonable model
for the emission? 
Both of them are answered affirmatively with 
the global fit of the data discussed later, based on \cite{new}.}
See \cite{malg,nadreview}
for further useful discussion of the time
sequence of the events.

\section{A two component model for $\bar\nu_e$ emission\label{sec3}}
Let us summarize the previous section.
There is some disturbing feature of the data.
The presence of few elastic scattering events does not seem to be 
essential to explain the angular distribution, 
that conforts us to proceed with the usual IBD 
hypothesis for the signal.
We showed that 
background events should be accounted for
and that it is important to consider very seriously 
the existence of an initial phase of high neutrino
luminosity. 
In short, the opinion that we derive from the previous discussion 
is that there is no serious problem with the data.
Thus we move to interpret them within the `standard scenario'.

\begin{figure}[t]
$$\includegraphics[width=.38\textwidth,
height=0.50\textwidth,
angle=270]{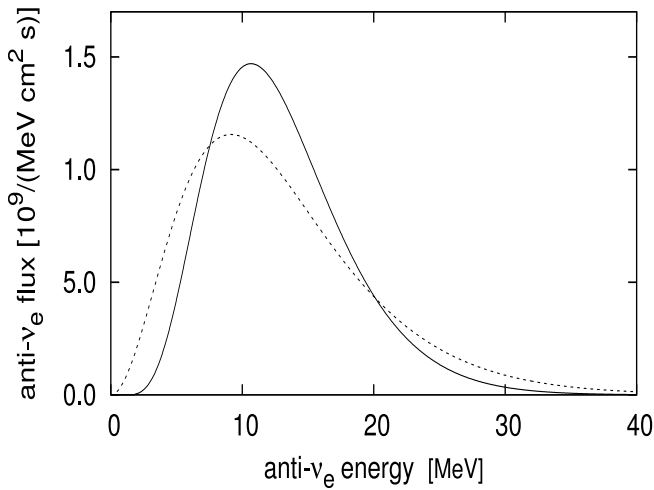}$$\vskip-4mm
\caption{\em \footnotesize 
Continuous curve: instantaneous 
$\bar\nu_e$ flux for $M_a=0.15\ M_\odot$ and $T_a=2.5$ MeV.
Dotted curve: black body distribution with the same luminosity
($1.1\times 10^{53}$~erg/s) and 
average energy ($13$~MeV), 
i.e., with parameters $R_c=82$ km and $T_c=4.1$~MeV. 
\label{fig:pinch}}
\end{figure}

We analyzed the 
data of Kamiokande-II, IMB and Baksan  
adopting a definite model for neutrino emission \cite{new}. 
The first attempt in this direction has been done by Lamb and 
Loredo~\cite{ll}.
Of course, the choice of the model is a critical step. 
We assume the existence of the two main 
phases of antineutrino emission as 
expected in the `standard scenario'. 
For each phase we need to describe the luminosity of the emission, 
the average antineutrino energy and the duration. 
For {\bf cooling}, the parameters  
are $R_c$, $T_c$ and $\tau_c$ 
(similar to the parameters of the previous section). For {\bf accretion},
we use the mass of neutrons exposed to the positrons, $M_a$, 
the temperature of the positrons, $T_a$, 
and the duration of the accretion, $\tau_a$. 
In fact, the $\bar\nu_e$ are produced 
from the reaction of the thermal 
positrons with the target neutrons around the proto neutron star,
through $e^+ n\to p\bar\nu_e$, the inverse of IBD; a sample energy 
spectrum is shown in Fig.~\ref{fig:pinch}.
Note the characteristic `pinching' of the distribution--an 
output, not an input of our model.

Let us mention for completeness some qualifying 
feature of our analysis.
We prescribe that 
(1)~the temperature of $e^+$ 
increases during accretion so that 
the average energy of $\bar\nu_e$ is 
approximatively continuous at $t\sim \tau_a$, that  overcomes 
the shortcoming of the Lamb and Loredo model noted in~\cite{mirizzi};
(2)~the number of neutrons exposed to 
the positron flux decreases in time more smoothly 
than as in \cite{ll};  
in this way the luminosity is also continuous, as expected on general
basis.
(3)~Finally we avoid the simultaneous presence of cooling and accretion 
$\bar\nu_e$ delaying  
the cooling phase by an amount $\sim \tau_a$ again 
improving on~\cite{ll}.
We also improved the energy spectrum 
of accretion neutrinos and included neutrino 
oscillations with normal mass hierarchy.

\begin{figure}[t]
$$\includegraphics[width=0.38\textwidth,
height=0.50\textwidth,
angle=270]{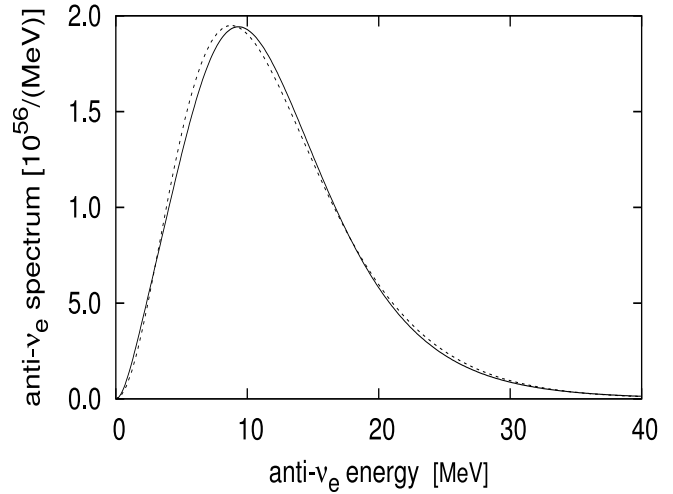}$$\vskip-4mm
\caption{\em \footnotesize 
The continuous curve is the $\bar\nu_e$ 
spectrum in the best fit model; 
the dotted curve is the approximant described in 
Eq.~\ref{appro}.\label{fig:fluk}}
\end{figure}

Our best fit result for the parameters is \cite{new}:
\begin{equation}
\begin{array}{ccc}
R_c=16\mbox{ km}, & T_c=4.6\mbox{ MeV}, & \tau_c=4.7\mbox{ s}, \\
M_a=0.22\ M_\odot, & T_a=2.4\mbox{ MeV}, & \tau_a=0.6\mbox{ s}.
\end{array}
\label{nb}
\end{equation}
The radiated binding energy is ${\cal E}_b=2.2\times 10^{53}$ erg, 
similar to the expected value for the formation  of a neutron star. 
The values of the parameters of the cooling phase 
are reasonable; in particular, $R_c$ resembles a
typical neutron star radius.
Also the values of the parameters of accretion resemble 
the expectations; the value of initial
accreting mass $M_a$ is a fraction of the outer core mass $\sim 0.6M_\odot$.
A last interesting outcome is that in the 30~s window 
Baksan had 0, 1, 2 background events 
with a posteriori probabilities of 
20\%, 47\%, 29\%, that compares well 
with the {\em a priori} expectation of 1 event.

The electron antineutrino spectrum, shown in 
Fig.~\ref{fig:fluk} can be calculated from the fluence 
(=time integrated flux) assuming $D=50$~kpc, and it 
is well approximated by a modified Fermi-Dirac
distribution: 
\begin{equation}
\frac{dN}{dE}=\frac{\kappa}{T^3} 
\frac{E^2}{1+\exp\left(\frac{E}{T}-\eta\right)} \mbox{ with }
\left\{
\begin{array}{l}
\kappa=9.45\times 10^{56}, \\
T=3.77\mbox{ MeV},\\
\eta=0.531. 
\end{array}
\right.\label{appro}
\end{equation} 
The {\em a posteriori} comparison with the detected energies 
is satisfactory, as already shown in Fig.~\ref{fig5}.

As a possible application, we use this result to 
predict the number of relic neutrino events.
We consider a detector a la SK 
with a fiducial mass of 22.5 kton \cite{malek}.
The rate of accumulation of 
IBD events is: 
\begin{equation}
   N_p \!\! \int_{\mbox{\tiny 19.3 MeV}}^{E_{max}} 
\!\!\!\! dE \sigma(E) 
\frac{c}{H_0} \! \int_{0}^{z_{max}} \!\!\! dz 
\frac{R(z)\ 
\displaystyle 
\frac{dN}{dE}((1+z) E)}{\sqrt{\Omega_\Lambda+\Omega_{DM}(1+z)^3}} 
\label{osti}
\end{equation}
The cosmological parameters are $H_0=70$ km/sec/Mpc, 
$\Omega_\Lambda=0.7$, $\Omega_{DM}=0.3$. 
An important quantity is the rate of core collapse 
supernovae $R$ as a   function of the redshift $z$, 
adequately described by: 
$R(z)=R(0) (1+z)^\beta$ if $z<1$ and 
$=R(0)2^\beta$ if $z\ge 1$.
$R(0)$ is the product of the fraction of 
core collapse supernovae $f_{SN}$ and 
the cosmic rate of star formation $R_*(0)$.
Other quantities 
(the maximum energy $E_\nu\sim 40-60$ MeV 
and redshift $\sim 5-6$, the point
$z=1$ where the slope is modified)
are not critical for the prediction.
We compare in Tab.~\ref{tab2} the events per year for three descriptions 
of the rate of cosmic supernovae. The lower number of events 
is rather close to the one given in~\cite{lunardini}; in other words, 
the non-standard neutrino spectrum used in~\cite{lunardini} gives
essentially the same results as the (more standard) spectrum of
Eq.~\ref{appro}.
Barring the possibility that SN1987A was a peculiar event, 
we are lead to believe that the largest uncertainty in predicting 
the relic neutrino signal is the rate of cosmic supernovae
rather than the model for neutrino emission.

\begin{table}[t]
\centerline{
\begin{tabular}{||c|c||cc|cc||c||}
\hline
$R(0)$  & $\beta$ & 
$N_{tot}$ & $\langle z\rangle$ &  $N_{thr}$ & $\langle
  z\rangle$
&  model \\
\hline 
$2.00$ & 2.00& 3 &0.5 & 0.6 &0.2 & ref.\cite{john}  \\
$0.67$ & 3.44& 2 &0.7& 0.3  &0.2 & ref.\cite{lunardini}  \\
$1.25$ & 3.44& 4 &0.7& 0.5  &0.2 & ref.\cite{john2} \\
\hline
\end{tabular}}
\caption{\em\footnotesize 
$1^{st}$ and $2^{nd}$ column, parameters of SN 
distribution--$R(0)$ is in units of 
$1/(10^{4}\mbox{Mpc}^{3} \mbox{yr})$; 
$3^{rd}$ and $4^{th}$ column, total number of expected events 
per year from the relic supernovae and average redshift;
$5^{th}$ and $6^{th}$ column, events above a threshold of 19.3 MeV
and average redshift.\label{tab2}}
\end{table}

\section{Discussion}

The difficulty to interpret SN1987A data is due to
small-number statistics, partial operativity of IMB, 
poor absolute times in Kamiokande-II and Baksan, 
an apparent excess of events in Baksan, 
``gap'' in the time-distribution of Kamiokande-II, 
peculiar angular distributions 
and different energies of Kamiokande-II and IMB. 
However, none of these problems
constitutes an unsurmountable difficulty 
and on the contrary they can and have to be taken
care in the analysis; furthermore, none of them
can be solved too directly by considering 
reasonable deviations from the standard expectations.
This does not guarantee us the possibility of reaching
a safe interpretation of the data, but contributes to make 
less arbitrary the belief that these unique data 
can be understood in a coherent framework. 

A  general question that we addressed is if the data resemble the 
conventional expectations. We motivated the opinion that 
the presence of background events should be accounted for and that 
the signal should  be described by considering an initial phase of
intense neutrino luminosity, as required in the `standard scenario'
\cite{nad,bw}.
This opinion is in agreement with the conclusions 
reached in 1989 in the review 
on SN1987A by Imshennik and Nadyozhin \cite{in}.

Being aware of the incompleteness of the present information
(on 3D effects, rotation, magnetic fields; on peculiarities of
SN1987A; on the existence of the neutron star; on neutrino
oscillations; on the detailed detector response, {\em etc} \cite{symp}) 
we attempted to define the expectations of the `standard scenario' 
\cite{nad,bw} in a
parameterized model, trying to keep it as simple and flexible as
possible but imposing certain general requirements,  
such as the continuity of the antineutrino luminosity and average energy.

We elaborated on the pioneer approach of Lamb and
Loredo to SN1987A data analysis \cite{ll} emphasizing its 
defects and drawbacks  but arguing that their model 
for antineutrino emission can be improved to 
resemble more closely the expectations \cite{new}.
The numerical analysis of the data 
yielded indications in (perhaps surprising) agreement with the 
expectations of the `standard scenario' for neutrino emission.
We estimated in this way 
the spectrum of $\bar\nu_e$ from the SN1987A data and 
applied this result to evaluate the number of 
relic neutrinos events.

\vskip2mm
\noindent 
{\scriptsize 
{\em Acknowledgments} F.V.\ thanks 
the Organizers of the Symposium for 
the honorable invitation; 
O.~Ryazhskaya for many stimulating discussions and 
inexhaustible passion with SN1987A; 
D.~Nadyozhin for help with 
astrophysics and precious encouragement; 
E.~Alexeev and A.~Mal'gin 
for frank and useful discussions 
on the meaning of the data;
and last but not least my friends 
N.~Agafonova, A.~Mal'gin, and V.~Yakushev
who made my first visit to Moscow unforgettable.   
\vskip-2mm}


\end{document}